\documentclass[aps,prb,reprint,superscriptaddress,longbibliography,floatfix]{revtex4-2}

\usepackage{amsmath,amssymb,bm}
\usepackage{graphicx}
\usepackage{placeins}
\usepackage[colorlinks=true,citecolor=blue,linkcolor=blue,urlcolor=blue]{hyperref}

\graphicspath{{figures/}}
\newcommand{\figcol}[1]{\centering\includegraphics[width=\columnwidth]{#1}}
\newcommand{\figtw}[1]{\centering\includegraphics[width=\textwidth]{#1}}
\newcommand{\wcav}{\omega_{\mathrm{cav}}}
\newcommand{\wB}{\omega_B}
\newcommand{\lcav}{\lambda_{\mathrm{cav}}}

\begin{document}

\title{Vacuum-field-mediated Landau-level reconstruction in a cavity-coupled two-dimensional electron gas}

\author{Zi-Heng Zhu}
\affiliation{Graduate School of China Academy of Engineering Physics, Beijing 100193, China}
\author{Yu Chen}
\thanks{Corresponding author.}
\affiliation{Graduate School of China Academy of Engineering Physics, Beijing 100193, China}

\begin{abstract}
We study a two-dimensional electron gas in the integer quantum Hall regime
coupled to a single spatially modulated cavity mode in the static
off-resonant limit.  Retaining both paramagnetic and diamagnetic
contributions on equal footing, we derive a gauge-complete single-particle
Hamiltonian in which virtual photon exchange generates a periodic kinetic
renormalization and a diamagnetic scalar potential.  Both terms transfer
momentum by $2k_0$ and displace guiding centers by
$\Delta X=2k_0\ell_B^2$, forming a finite-$k_0$ momentum ladder.
Diagonalizing the strip Hamiltonian, we find that the ladder reconstructs
the Landau-level band structure and redistributes, rather than multiplies,
the associated chiral edge states.  Occupied-subspace Chern numbers remain
$C_\nu=\nu$ for $\nu=1,2,3$ throughout $0\le\lambda_{\mathrm{cav}}\le 0.5$, while the bulk
gaps decrease without closing.  Strip Kubo responses show cavity-induced
displacements of the Hall-step energies; at finite width the plateau
heights are noninteger and approach the integers as $L_x$ increases.  The
ladder is spectrally resolved at $k_0\ell_B=0.5$ and collapses at
$k_0\ell_B=0.1$.  These results isolate a spatially local, finite-$k_0$
momentum ladder as a clean-system mechanism by which a structured vacuum
field reconstructs quantum Hall spectra and strip transport without
changing the integer topology, complementary to cavity-mediated all-to-all
hopping.
\end{abstract}

\maketitle

\section{Introduction}

The integer quantum Hall effect (IQHE) is the quintessential example of
topology in physics~\cite{vonKlitzing1980,Laughlin1981,Buttiker1988}.
The robustness of its Hall plateaus is rooted in the interplay between
bulk topology and Anderson localization induced by
disorder~\cite{Pruisken1984,Khmelnitskii1984,Wei1988,Shahar1995,Ostrovsky2007PRL,Altland2014}.
Yet disorder is not only a localization mechanism: it can also actively
generate topology.  In topological Anderson insulators, random potentials
renormalize the Dirac mass and drive a trivial band insulator into a
topological phase~\cite{Groth2009,Li2009}.  A distinct realization of this
idea occurs in quasiperiodic quantum kicked rotors, where deterministic
chaos produces an effective disorder and gives rise to topologically
quantized insulating
phases~\cite{vanNieuwenburg2012,Edge2012,Chen2014,Tian2016,Titum2015,Titum2016,Kim2020AFAI}.
Taken together, disorder and chaos provide intrinsic routes to topological
phases, and a complementary question is whether topology can also be
influenced by quantum fluctuations of the electromagnetic vacuum.

This route is enabled by deeply subwavelength terahertz resonators, which
open a complementary direction by modifying a two-dimensional electron gas
(2DEG) with a structured electromagnetic vacuum.  Complementary split-ring
resonators concentrate near fields at the 2DEG and enable ultrastrong
coupling to cyclotron transitions, including Landau
polaritons~\cite{Hagenmuller2010,Li2018}.  Magnetotransport in this setting
is vacuum-field dependent~\cite{ParaviciniBagliani2019}.  Experiments have
reported a weakening of integer quantum Hall
protection~\cite{Appugliese2022}, a tuning of integer and fractional
gaps~\cite{Enkner2025}, and control of quantum Hall stripe
order~\cite{Graziotto2026}.  These observations motivate theoretical work
in which both the coupling strength and the spatial profile of the cavity
mode are control parameters.  They do not, by themselves, identify a
unique microscopic mechanism.

To interpret these observations, several theoretical routes have been
developed.  Uniform or collective cavity coupling yields anisotropic
magnetotransport~\cite{Bartolo2018} and polaritonic reconstructions of
Hofstadter spectra and Hall plateaus~\cite{Rokaj2022}.  Cavity-dressed
fractional quantum Hall liquids have also been
analyzed~\cite{Bacciconi2025}.  A prominent interpretation of the integer
quantum Hall transport experiments is cavity-mediated long-range hopping
generated by antiresonant light--matter processes that connect distant
electronic states through a shared
mode~\cite{Ciuti2021,Appugliese2022}.  That hopping range is set by the
cavity mode volume rather than by a single in-plane wave vector.

Here we isolate a different, spatially resolved single-particle mechanism.
We retain one cavity mode polarized along $\hat{\bm x}$ and modulated along
$\hat{\bm y}$ with wave vector $k_0$.  In the static off-resonant limit,
virtual photons generate a periodic kinetic renormalization together with
a diamagnetic potential term.  Both contributions contain Fourier
components $e^{\pm i2k_0 y}$ and therefore couple momenta
$k\leftrightarrow k\pm 2k_0$.  The associated guiding-center displacements
form a momentum ladder of step $\Delta X=2k_0\ell_B^2$.  This finite-$k_0$
ladder is complementary to cavity-mediated all-to-all hopping: it is local
on the scale of $\Delta X$, and it is realized here in a clean static
Hamiltonian rather than in a disordered polaritonic model.  Thus, the
present mechanism is distinct from both uniform-cavity anisotropy and
cavity-mediated all-to-all hopping.

For comparison, near-resonant experiments that weaken integer quantum Hall
protection also rely on disorder~\cite{Appugliese2022,Enkner2025}.
Plateau erosion then tracks all-to-all hopping when
$\Omega_R\simeq 0.19\,\mathrm{meV}$ is comparable to
$\Gamma_{\mathrm{dis}}\sim 0.1\text{--}0.5\,\mathrm{meV}$~\cite{Ciuti2021,Appugliese2022}.
Equation~\eqref{eq:heff} retains only local Fourier sectors
$e^{\pm i2nk_0 y}$ and lacks that nonlocal vertex.  The present spinless
model is therefore not intended to describe the Zeeman-limited $\nu=1$
plateau; rather, it targets the cyclotron-gap-dominated $\nu=2$
regime~\cite{Appugliese2022}.

We study the clean static model at $r_\omega\equiv\wcav/\wB=5$ and
$k_0\ell_B=0.5$.  This baseline separates spectral reconstruction from
disorder and resonant polariton formation.  The remainder of the paper is
organized as follows.  Section~II introduces the effective model.
Section~III discusses the reconstructed spectrum and edge states.
Section~IV examines the occupied-subspace topology.  Section~V presents
the Kubo response.  Section~VI concludes.

\section{Effective model}

Consider a 2DEG in a uniform magnetic field $\bm B=B\hat{\bm z}$.  The
kinetic Hamiltonian is
\begin{equation}
  H_0=\frac{(\bm p+e\bm A_B)^2}{2m}.
\end{equation}
In the Landau gauge $\bm A_B=Bx\hat{\bm y}$ the system is translationally
invariant along $y$, so that $p_y=k$ is conserved.  $H_0$ reduces to a
one-dimensional harmonic oscillator centered at the guiding center
$X=-k\ell_B^2$, with magnetic length $\ell_B=(eB)^{-1/2}$.  The eigenstates
are
\begin{equation}
 \psi_{nk}(x,y)=\phi_n(x + k \ell_B^2)e^{iky},
 \label{eq:landau-wavefunction}
\end{equation}
with energies $E_n=(n+1/2)\wB$ and cyclotron frequency $\wB=eB/m$.

\begin{figure}[t]
 \figcol{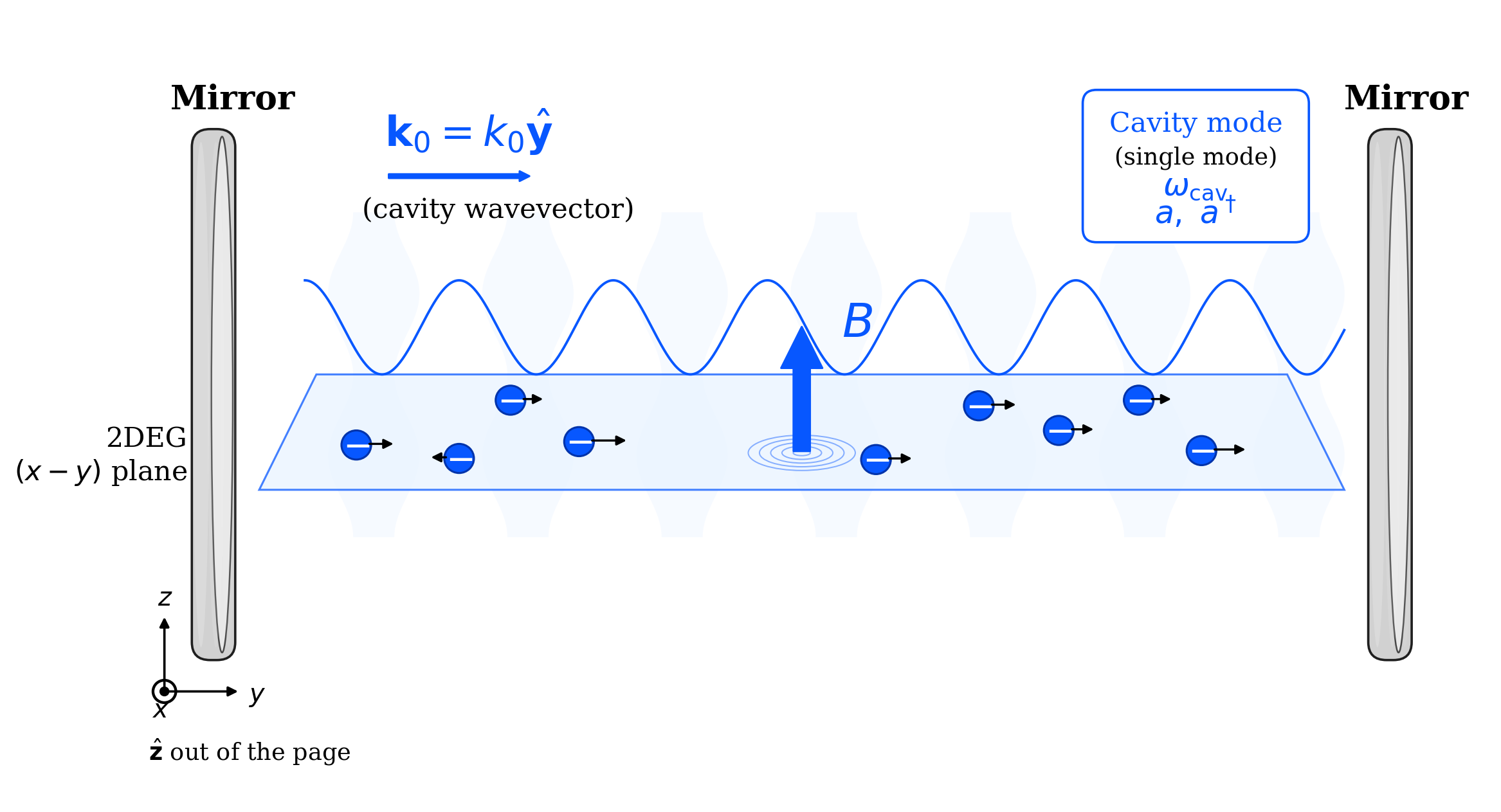}
 \caption{Cavity-coupled Landau-level geometry.  A 2DEG in the $x$--$y$
 plane is subject to a perpendicular field $B\hat{\bm z}$.  The selected
 cavity mode is polarized along $\hat{\bm x}$ and spatially modulated along
 $\hat{\bm y}$ with wave vector $k_0$.  Spectral and transport calculations
 use a strip with hard walls at $x=\pm L_x/2$ and periodic boundary
 conditions along $y$; chiral edge states therefore propagate along $y$.
 The mirrors are schematic.}
 \label{fig:setup}
\end{figure}

The 2DEG is placed in the cavity geometry of Fig.~\ref{fig:setup}.  We
retain a single mode polarized along $x$ and modulated along $y$, with
vacuum vector potential
$\bm A_{\mathrm{cav}}(y)=A_0\cos(k_0y)(a+a^\dagger)\hat{\bm x}$.
$A_0$ is the electron-cavity coupling strength.  Hard-wall
confinement, used later for the strip spectra, is along $x$; the
propagation direction of the chiral edge states is then $y$.

Minimal coupling $\bm p\to\bm p+e\bm A_B+e\bm A_{\mathrm{cav}}$ gives
\begin{align}
 H={}&H_0+\wcav a^\dagger a+V_{\mathrm p}+V_{\mathrm d},
                                                               \label{eq:fullH}\\
 V_{\mathrm p}={}&\frac{eA_0}{m}\cos(k_0y)(a+a^\dagger)p_x,            \label{eq:vp}\\
 V_{\mathrm d}={}&\frac{e^2A_0^2}{2m}\cos^2(k_0y)
 (a+a^\dagger)^2.                                             \label{eq:vd}
\end{align}
The paramagnetic
vertex therefore connects Landau states with momenta $k$ and $k\pm k_0$,
while $V_{\mathrm d}$ is diagonal in the photon-number basis up to
two-photon terms.

We eliminate photonic excitations in the vacuum sector.  Let
$P=|0\rangle\langle0|$ and $Q=1-P$.  Because $PV_{\mathrm p}P=0$, the
leading vacuum correction is second order in $A_0$,
\begin{equation}
 H_{\mathrm{eff}}(E)=H_0+PV_{\mathrm d}P
 -PV_{\mathrm p}Q \frac{1}{\wcav a^\dagger a-E+H_0}Q V_{\mathrm p}P.
 \label{eq:projection}
\end{equation}
In the static off-resonant approximation we replace the resolvent by
$1/\wcav$ and retain both $O(A_0^2)$ contributions. Defining
\begin{equation}
 \lcav=\frac{2e^2A_0^2}{m\wcav},
 \qquad
 U_d=\frac{e^2A_0^2}{2m}
 =\frac{\lcav\wcav}{4}
 \label{eq:lambda-ud}
\end{equation}
yields the gauge-complete static Hamiltonian used throughout,
\begin{equation}
 H_{\mathrm{eff}}\simeq
 \frac{p_x^2}{2m}\left[1-\lcav
 \cos^2(k_0y)\right]+\frac{\pi_y^2}{2m}
 +U_d\cos^2(k_0y).                                     \label{eq:heff}
\end{equation}
Equation~\eqref{eq:heff} retains both the diamagnetic $A^2$ and
paramagnetic $\bm A\cdot\bm p$ corrections on equal footing, so
observables such as the full strip $\sigma_{xx}$ and the Chern markers
remain gauge-invariant.  This contrasts with phenomenological models that
kept only the paramagnetic vertex.  Appendix~\ref{app:pf} compares
Eq.~\eqref{eq:heff} with the parent Pauli--Fierz strip Hamiltonian: the
static model reproduces the low-energy spectrum throughout
$0\le\lcav\le 0.5$, whereas dropping $A^2$ leaves a systematic error that
does not vanish at large $\wcav$.
The kinetic term produces a spatially modulated anisotropic mass.  The
scalar $U_d$ term shifts and mixes the same Fourier sectors without changing
the local orbit aspect ratio.  Expanding
$\cos^2(k_0y)=\tfrac12+\tfrac14(e^{i2k_0y}+e^{-i2k_0y})$ shows that both
corrections transfer momentum by $\pm 2k_0$ and displace guiding centers by
$\Delta X=2k_0\ell_B^2$.

All numerical scans fix $r_\omega\equiv\wcav/\wB=5$, so that
$U_d/\wB=\lcav r_\omega/4$ is locked to $\lcav$ and is not varied
independently.  Positivity of the local kinetic coefficient requires
$\lcav<1$.  We restrict the scans to $0\leq\lcav\leq 0.5$, for which that
coefficient remains at least one half.  This interval is a weak-to-moderate
window of the truncated static model, not a claim of validity throughout
the ultrastrong-coupling regime.

A useful scale for the $2k_0$ hybridization follows by projecting
Eq.~\eqref{eq:heff} onto the lowest Landau level.  The leading hop between
guiding centers separated by $\Delta X$ is
\begin{equation}
 t_1=\frac{\lcav\wB}{16}(r_\omega-1)\,e^{-(k_0\ell_B)^2}.
 \label{eq:t1}
\end{equation}
At the working value $r_\omega=5$ the diamagnetic piece dominates and fixes
the sign of $t_1$.  Higher harmonics that transfer $4k_0$, $6k_0$, \ldots\
are higher order in $\lcav$ or involve virtual Landau-level excitations.
Equation~\eqref{eq:t1} is an estimate of the ladder scale; the spectra and
conductivities below are obtained from the continuum strip Hamiltonian
rather than from a truncated lowest-Landau-level network.

The dimensionless product $k_0\ell_B$ controls the real-space range of the
ladder.  Deeply subwavelength microwave and terahertz resonators typically
realize $k_0\ell_B\ll 1$, for which $\Delta X/\ell_B=2k_0\ell_B$ is small and
the reconstruction is weak.  The strip calculations use $k_0\ell_B=0.5$,
so that several
Fourier sectors fit across $L_x=20\ell_B$ and the ladder is spectrally
visible.  That choice
is a numerical working point, not a claim that present terahertz devices
operate in the same regime.
Appendix~\ref{app:sec4_k0} confirms that expectation at
fixed $\lcav=0.2$: the $2k_0$ copies overlap at $k_0\ell_B=0.1$ and become
spectrally resolved at $k_0\ell_B=0.5$.

\section{Spectrum and edge-state reconstruction}
\label{sec:spectrum}

We diagonalize Eq.~\eqref{eq:heff} on a hard-wall strip
$x\in[-L_x/2,L_x/2]$ with periodic boundary conditions along $y$.  The
reduced quasi-momentum is $q\in[-k_0,k_0)$, and the Fourier sectors carry
physical momenta $k_n=q+2nk_0$.  Unless noted otherwise we use units
$\ell_B=\hbar=\wB=1$.
Spectral figures plot $E-E_{\min}+1/2$; fillings and Kubo scans
use the same shift with the Fermi energy $E_F$.
The spectral mesh is $L_x=20$, $N_x=240$, $k_0=0.5$,
$r_\omega=5$, $N_{\mathrm{modes}}=21$, and $N_q=41$, retaining the lowest
$120$ eigenstates at each $q$.  The highest retained state lies above
$5.41$, so the displayed window below $3.2$ is complete.
Transport calculations below use the denser mesh $N_x=360$,
$N_{\mathrm{modes}}=25$, and $N_q=81$.
The strip width $L_x=20\ell_B$ is a numerical working point chosen so that
several displacements $\Delta X=2k_0\ell_B^2$ fit across the strip and the
ladder hybridization is spectrally visible; finite-width transport is
deferred to Appendix~\ref{app:lx}.

\begin{figure*}[t]
 \figtw{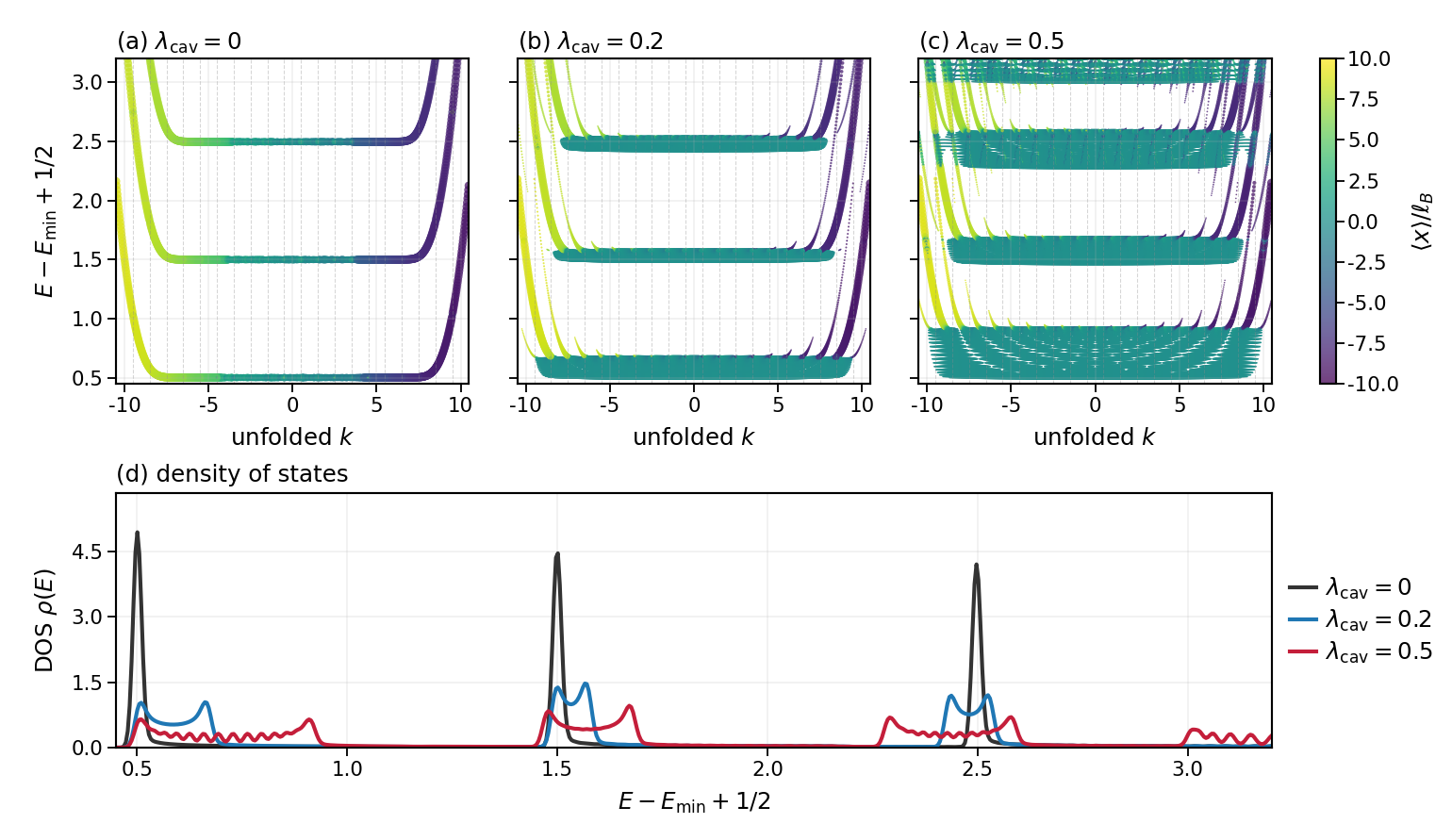}
 \caption{Spectral reconstruction at representative couplings.
 Panels (a)--(c): unfolded spectral weights for
 $\lcav=0$, $0.2$, and $0.5$.  Point size encodes the Fourier-sector
 weight of each eigenstate and color encodes
 $\langle x\rangle/\ell_B$.  Repeated Landau-level branches at momenta
 shifted by $2nk_0$ are projections of the same reduced-zone eigenstate,
 not additional independent edge branches.
 Panel (d): strip density of states for the same three couplings, obtained
 with Gaussian broadening $\Gamma=0.01\wB$.}
 \label{fig:spectral}
\end{figure*}

Figure~\ref{fig:spectral}(a)--(c) shows the reconstruction in the unfolded
zone.  At $\lcav=0$ the spectrum consists of flat bulk Landau levels and
dispersive edge branches, each carrying a single dominant Fourier
component.  Finite coupling generates replicas at $k\pm 2k_0$, $k\pm 4k_0$,
\ldots.  These replicas are the same physical state projected onto different
momentum sectors: their real-space centers coincide, and their weights
diminish with harmonic order.  The reconstruction is therefore not a rigid
energy shift.  Because the coupling matrix elements depend on the
Landau-level and guiding-center components, the momentum-ladder sectors
hybridize nonuniformly, producing avoided crossings and enhanced subband
dispersion in the higher Landau manifolds.

The same reconstruction is controlled by $k_0$.
Appendix~\ref{app:sec4_k0} shows the copies collapsing onto
one Landau-level branch at $k_0\ell_B=0.1$, and the hybridized subbands of
Fig.~\ref{fig:spectral}(b) at $k_0\ell_B=0.5$.

\begin{figure*}[t]
 \figtw{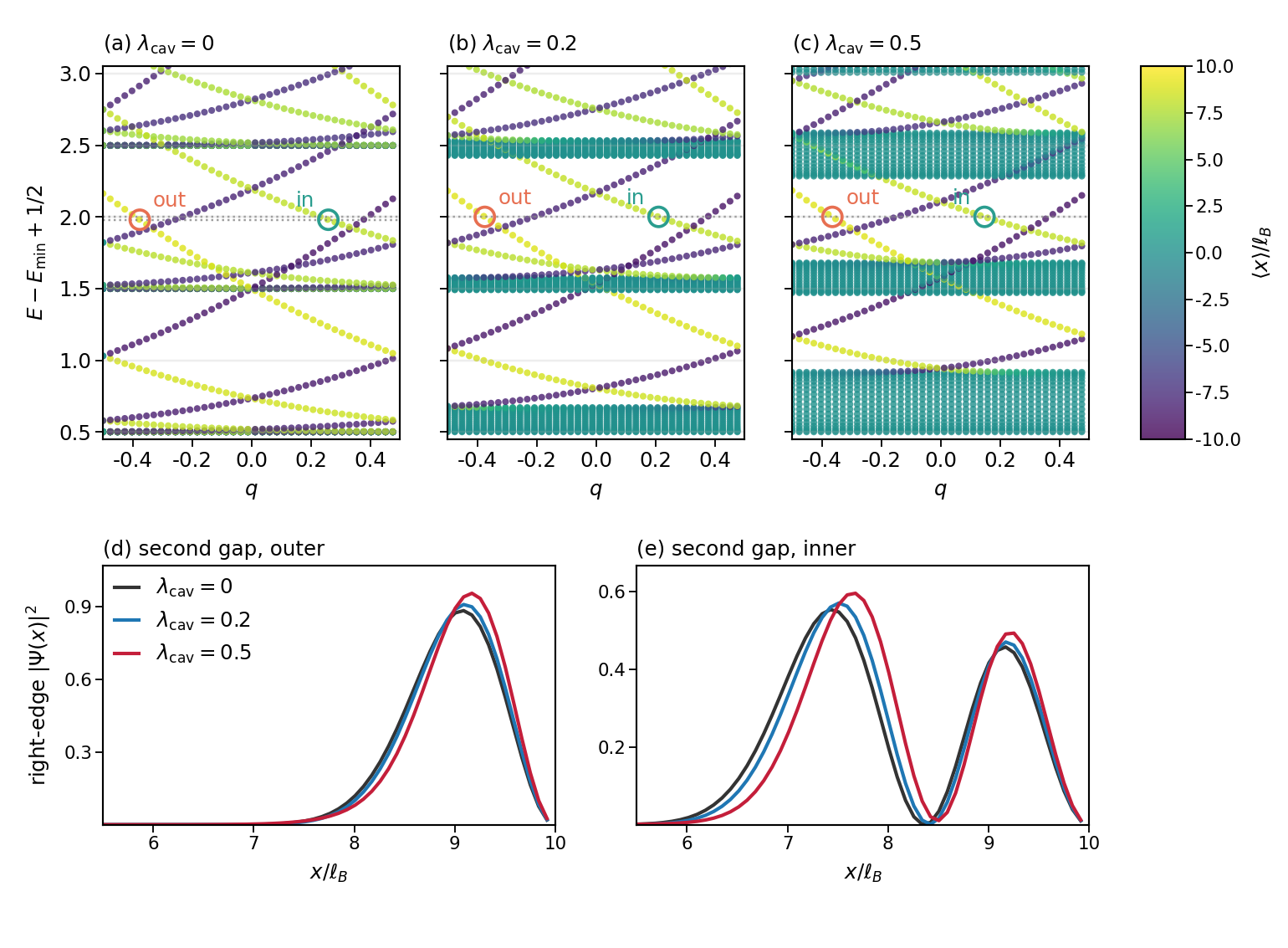}
 \caption{Second-gap reduced-zone spectra and right-edge profiles.
 Panels (a)--(c): reduced-zone dispersions at
 $E-E_{\min}+1/2=2$ for $\lcav=0$, $0.2$, and $0.5$.  Orange and green
 circles mark the outer and inner right-edge branches used below.
 Panels (d)--(e): right-edge $|\Psi(x)|^2$ at $x\in[5.5,10]\ell_B$ for
 those two branches, with $\lcav=0$ (black), $0.2$ (blue), and $0.5$
 (red).}
 \label{fig:edges}
\end{figure*}

The reduced-zone representation in Fig.~\ref{fig:edges} avoids double
counting.  At
$E_F-E_{\min}+1/2=1$
the uncoupled and cavity-coupled strips
each contain one left--right chiral pair; at
$E_F-E_{\min}+1/2=2$
each
contains two.
Figure~\ref{fig:edges}(a)--(c) focuses on the second gap and marks one
outer and one inner right-edge branch at each coupling.  Cavity coupling
reconstructs the dispersions without changing the two-pair count.
Additional unfolded replicas are therefore Fourier copies of the same
physical branches, not new independent edge channels.

The comparison panels in Fig.~\ref{fig:edges}(d)--(e) show that these
right-edge profiles remain wall-bound on the displayed scale.  At
$\lcav=0.2$ they nearly coincide with the uncoupled curves; only at
$\lcav=0.5$ does the outer branch develop a slightly sharper peak and
the inner branch a modest inward shift.  Distinct secondary maxima
separated by $\Delta X$ are not resolved.  The reconstruction that is
unambiguous is therefore the Fourier-sector hybridization of
Fig.~\ref{fig:spectral} and the reduced-zone dispersions of
Fig.~\ref{fig:edges}, not a bulk-penetrating rearrangement of
$|\Psi(x)|^2$.

The lower panel of Fig.~\ref{fig:spectral} summarizes the spectral
evolution.
The second and third Landau peaks move downward and broaden
with $\lcav$, consistent with the avoided crossings of
Fig.~\ref{fig:spectral}(a)--(c).  The lowest peak is more rigid.  These are
clean single-particle features of Eq.~\eqref{eq:heff}.  They indicate that
a finite density of states appears over a wider energy window, but they do
not establish a mobility edge or an impurity scattering rate.

\section{Bulk topology}
\label{sec:topology}

The reconstructed spectrum and the unchanged reduced-zone channel count
leave open whether the occupied-subspace topology itself is preserved.  We
place a periodic square-lattice regularization of Eq.~\eqref{eq:heff} on a
torus and impose boundary twists $(\theta_x,\theta_y)$.  The magnetic field
is implemented by Peierls phases.  The cavity modifies the real
$x$-direction hopping,
$t_x(j)=[1-\lcav\cos^2(k_0ja)]/(2a^2)$,
and adds the on-site diamagnetic potential
$U_d^{\mathrm{lat}}\cos^2(k_0ja)$ with
$U_d^{\mathrm{lat}}=(\lcav r_\omega/4)\,(2\pi\phi)$
in lattice units at flux $\phi$ per plaquette.  Hopping along $y$ remains
uniform.  For all states below a selected bulk gap we evaluate the
gauge-invariant non-Abelian Fukui--Hatsugai--Suzuki sum over twist
space~\cite{Fukui2005}.  This gives the total Chern number of the occupied
subspace and remains well defined in the presence of degeneracies within
that subspace.

\begin{figure}[t]
 \figcol{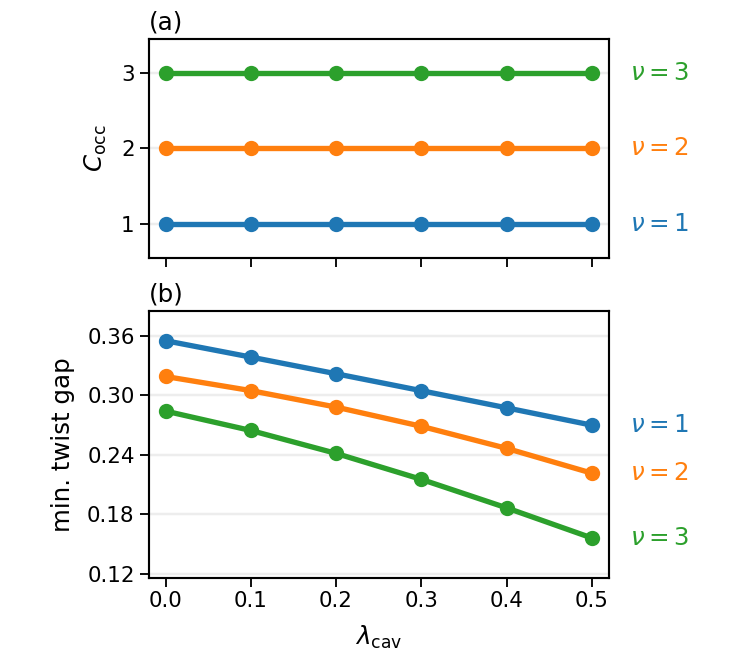}
 \caption{Occupied-subspace Chern numbers $C_{\mathrm{occ}}$ (a) and
 minimum direct gaps over the twist grid (b) versus $\lcav$.  The
 calculation uses $N_x\times N_y=16\times 8$, flux $\phi=1/16$ per
 plaquette, $k_0a=\pi/4$, $r_\omega=5$, and a $15\times 15$ twist grid.
 The first three Chern numbers remain $C_\nu=\nu$.  The gaps decrease
 monotonically and none closes.  A $21\times 21$ twist grid yields the same
 integers.}
 \label{fig:chern}
\end{figure}

Figure~\ref{fig:chern} gives
\begin{equation}
 C_{\nu}(\lcav)=C_{\nu}(0)=\nu,
 \qquad \nu=1,2,3,
 \label{eq:chern-stability}
\end{equation}
throughout $0\leq\lcav\leq 0.5$.  The minimum direct gaps over the twist grid decrease
from approximately $0.355$, $0.320$, and $0.285$ at $\lcav=0$ to
$0.270$, $0.222$, and $0.157$ at $\lcav=0.5$, in lattice units.  None
closes.  A lower-flux check at $\phi=1/24$ likewise yields $C_3=3$ up to
the largest coupling.
The gap reduction tracks the downward shift and broadening of the higher
Landau peaks in Fig.~\ref{fig:spectral}(d).  Because the occupied subspace
remains gapped, this quantization is the expected rigidity of a continuous
deformation of a Chern insulator.  It confirms the reduced-zone channel
counts of Fig.~\ref{fig:edges}.  It does not by itself address disordered
mobility gaps or experimental plateau widths.

\section{Kubo responses}
\label{sec:kubo}

Because the strip is clean, the diagonal Drude weight is either
divergent ($\propto 1/\eta$) or purely $\eta$-dependent.  We therefore
report the interband spectral weight of Eq.~\eqref{eq:kubo-long}.
On the hard-wall strip, $v_{nn}^{x}\equiv 0$, so
$\sigma_{xx}^{\mathrm{inter}}$ is the complete transverse dissipative
response.  Along the edge,
$v_{nn}^{y}=\partial E_n/\partial q\neq 0$ on chiral branches, and the
same formula yields only a truncated remainder of $\sigma_{yy}$.
Inequality of the two traces, even at $\lcav=0$, reflects that
truncation and the strip geometry, not a bulk DC anisotropy.

Using the same strip eigenstates, we evaluate the zero-temperature Kubo
conductivities with Lorentzian broadening $\eta$.  The Hall response is
\begin{equation}
 \sigma_{xy}=\frac{i}{A}\sum_{m\neq n}
 \frac{\bigl(f(E_m)-f(E_n)\bigr)
 \langle m|v_x|n\rangle\langle n|v_y|m\rangle}
 {(E_m-E_n)(E_m-E_n+i\eta)},
 \label{eq:kubo-xy}
\end{equation}
and the
interband longitudinal weights
\begin{equation}
 \sigma_{\mu\mu}^{\mathrm{inter}}=\frac{\pi}{A}\sum_{m\neq n}
 \frac{f(E_m)-f(E_n)}{E_m-E_n}
 \frac{(\eta/\pi)\bigl|\langle m|v_\mu|n\rangle\bigr|^2}
 {(E_m-E_n)^2+\eta^2},
 \label{eq:kubo-long}
\end{equation}
with area $A$ and Fermi--Dirac factors $f$ at $T=0$.  Conductivities are
reported in units of $e^2/h$.  The velocity operators follow from
Eq.~\eqref{eq:heff},
\begin{equation}
 v_x=\frac{p_x}{m}\bigl[1-\lcav\cos^2(k_0 y)\bigr],
 \qquad
 v_y=\frac{\pi_y}{m}.
 \label{eq:velocities}
\end{equation}
The scalar $U_d$ term does not contribute to $\bm v$.  The cavity therefore
modulates only $v_x$, while $v_y$ remains the along-edge operator of the
Landau-gauge strip.
The main transport scan uses $L_x=20\ell_B$, $N_x=360$,
$N_{\mathrm{modes}}=25$, $N_q=81$, and $\eta=0.01\wB$.  This broadening is
small compared with the cyclotron spacing and is used to smooth the discrete
strip spectrum.  It is not an impurity self-energy: the Hamiltonian in this
section remains clean.

\begin{figure}[t]
 \figcol{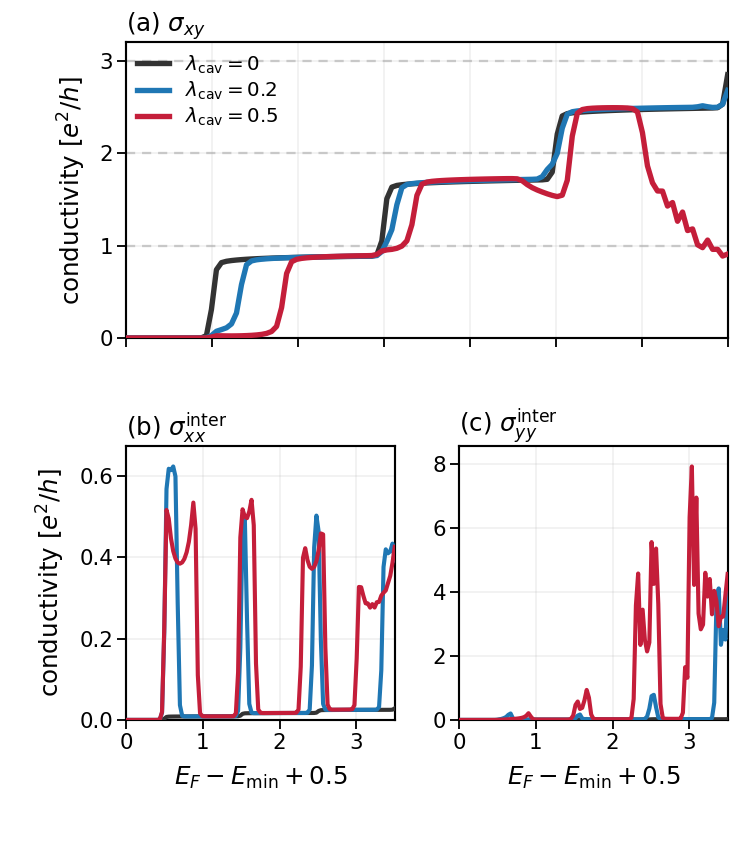}
 \caption{Kubo conductivities versus band-minimum-referenced Fermi energy
 for $\lcav=0$, $0.2$, and $0.5$ at $L_x=20\ell_B$ and $\eta=0.01\wB$.
 Top: Hall conductivity $\sigma_{xy}$.  Bottom left: transverse
 $\sigma_{xx}^{\mathrm{inter}}$, which is the complete $\sigma_{xx}$ of
 the hard-wall strip.  Bottom right: along-edge
 $\sigma_{yy}^{\mathrm{inter}}$, which omits the diagonal Drude weight.
 Hall steps shift with
 coupling, while the finite-width plateau values remain noninteger.
 }
 \label{fig:conductivity}
\end{figure}

Figure~\ref{fig:conductivity} collects the $\lcav$ scan.  Spectral
reconstruction moves the
Hall-step energies.
On the common band-minimum
reference, the second transition
shifts from about
$1.50$ at zero coupling to $1.375$ at $\lcav=0.5$, matching
the compressed Landau
spacing in the density of states.
The kinetic
modulation and the diamagnetic scalar $U_d$ both contribute; with
$r_\omega$ fixed they are not separated here.  The traces are step-like
but do not sit on integers: at $L_x=20\ell_B$ the first three apparent
plateau heights lie near $0.85$, $1.7$, and $2.5$.
Those noninteger heights are already present at $\lcav=0$.
Appendix~\ref{app:lx} shows that both uncoupled and moderately coupled
responses approach the integers as $L_x$ increases, while the
cavity-induced step displacements persist.
At the largest
coupling the third step distorts, in line with the
reconstructed high-energy spectrum rather
than a topological transition on the torus.

The two longitudinal
traces respond differently.  Peaks in $\sigma_{xx}^{\mathrm{inter}}$ remain
of order $0.5\,e^2/h$, whereas $\sigma_{yy}^{\mathrm{inter}}$ develops
larger high-energy peaks.  This records interband matrix elements on the
chosen strip: $v_y$ probes the reconstructed along-edge dispersion, while
$v_x$ is the cavity-modulated transverse operator.
Neither curve is a new impurity-scattering channel, nor a
Hall-bar resistance.

A calculation with the along-edge Drude weight, disorder, and
leads would still need to discriminate the present finite-$k_0$ ladder
from uniform-cavity anisotropy~\cite{Bartolo2018} and from
cavity-mediated long-range hopping~\cite{Ciuti2021}.
That nonlocal vertex is absent from Eq.~\eqref{eq:heff} and
becomes transport-relevant with disorder when
$\Omega_R\sim\Gamma_{\mathrm{dis}}$~\cite{Appugliese2022}.
A quantitative comparison with present terahertz devices remains outside
this off-resonant, clean calculation.

\section{Conclusion}
A spatially modulated cavity vacuum field couples quantum Hall states
through a momentum ladder of step $2k_0$.  Within the gauge-complete static
single-particle model, that ladder reconstructs Landau subbands and
redistributes, rather than multiplies, physical edge channels.
Occupied-subspace Chern numbers remain integer on a
clean torus, while bulk gaps decrease without closing.  On a finite strip
the same reconstruction shifts Hall-step energies.
Noninteger plateau heights at the working width are a finite-size effect,
not a change of topology.

These statements are confined to a detuned, clean, single-particle
Hamiltonian.  They do not establish an impurity-driven breakdown of edge
protection, nor do they provide a unified account of existing cavity
quantum Hall experiments.
The ladder is resolved at $k_0\ell_B=0.5$ and collapses at
$k_0\ell_B=0.1$.  Equation~\eqref{eq:heff} is the controlled off-resonant
limit of the parent Pauli--Fierz model in this window.
Remaining
uncertainties are mobility gaps under disorder, the near-resonant polariton
regime, and a comparison with cavity-mediated long-range hopping.
In that broader setting the finite-$k_0$ ladder remains
a distinct handle.  It ties spectral reconstruction and
direction-dependent matrix elements to the spatial structure of the vacuum
field, without a change of integer topology.

\begin{acknowledgments}
This work is supported by the National Key R\&D Program of China
(Grant No. 2022YFA1405302), the National Natural Science Foundation of
China (Grants No. 12675031 and No. 12174358), and NSAF
(Grant No. U2330401).
\end{acknowledgments}

\clearpage
\appendix
\section{Static approximation versus Pauli--Fierz}
\label{app:pf}
\begin{figure}[!ht]
 \figcol{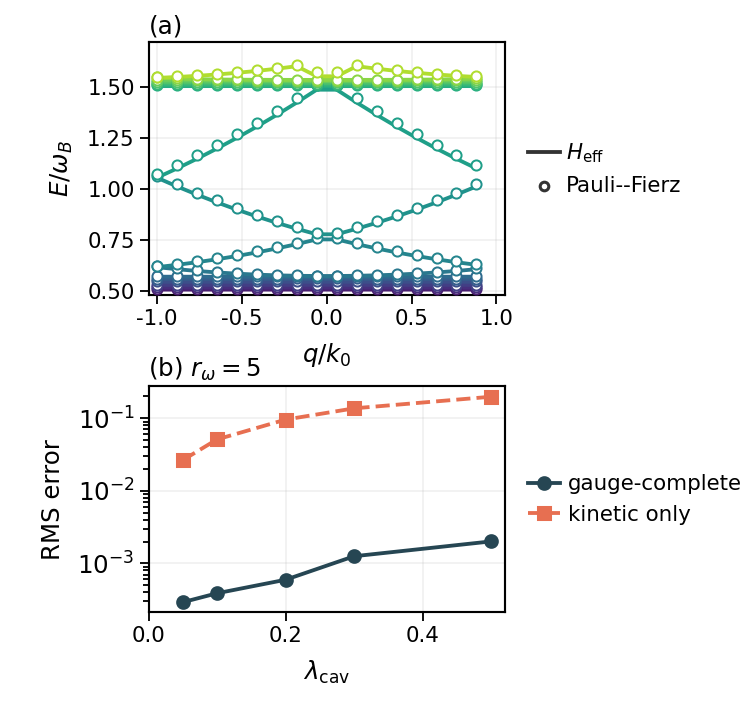}
 \caption{Pauli--Fierz benchmark of the static models at $k_0\ell_B=0.5$ and
 $r_\omega=5$, on a strip of width $L_x=10\ell_B$.
 (a)~Lowest sixteen bands: parent Pauli--Fierz Hamiltonian (open circles)
 and Eq.~\eqref{eq:heff} (solid).
 (b)~RMS excitation-energy error versus $\lcav$ at the working frequency
 ratio, evaluated on the lowest six bands for the gauge-complete static
 model (solid) and a kinetic-only truncation that omits $U_d$ (dashed).}
 \label{fig:pf}
\end{figure}

Equation~\eqref{eq:heff} is the static, off-resonant reduction of
the single-mode Pauli--Fierz strip Hamiltonian obtained from
Eqs.~\eqref{eq:fullH}--\eqref{eq:vd} by eliminating photonic excitations
in the vacuum sector.  Figure~\ref{fig:pf} tests that reduction on a
narrower strip ($L_x=10\ell_B$) that can accommodate an explicit Fock
cutoff.  The parent Hamiltonian is diagonalized in a hard-wall sine basis
with photon numbers $n_{\mathrm{ph}}=0,\ldots,4$ and momentum sectors
$|j|\le 10$, at $k_0\ell_B=0.5$ and $r_\omega=5$.  Eigenstates of
Eq.~\eqref{eq:heff} are matched to Pauli--Fierz states through the overlap
of the static wave function with the zero-photon projection, rather than
by energy ordering alone.

The leading virtual process generated by $V_{\mathrm p}$ connects
$|n,0\rangle$ to $|n\pm1,1\rangle$.  For a parabolic two-dimensional
electron gas the associated detuning is $\wB(r_\omega\mp 1)$ and does not
depend on the Landau index.  The static replacement of the resolvent by
$1/\wcav$ is therefore controlled by $r_\omega$, not by whether the
reconstructed manifold is the lowest or a higher Landau level.

At the baseline vacuum-field amplitude $\eta_A=eA_0\ell_B=0.45$
($\lcav=2\eta_A^2/r_\omega=0.081$), the gauge-complete static bands overlie
the Pauli--Fierz spectrum [Fig.~\ref{fig:pf}(a)].  The mean zero-photon
weight is $0.997$ and the mean photon number is
$2.92\times10^{-3}$.  Matched excitation-energy errors for the lowest
sixteen bands have RMS $5.51\times10^{-3}\,\wB$ for
Eq.~\eqref{eq:heff} and $4.23\times10^{-2}\,\wB$ if the diamagnetic
scalar $U_d$ is dropped.  A scan at the working
frequency ratio $r_\omega=5$, now using the lowest six bands, keeps the
gauge-complete RMS below $2.1\times10^{-3}\,\wB$ throughout
$0.05\le\lcav\le 0.5$, while the kinetic-only truncation grows from
$2.6\times10^{-2}\,\wB$ to $2.0\times10^{-1}\,\wB$
[Fig.~\ref{fig:pf}(b)].  The static Hamiltonian used in the main text is
therefore the controlled off-resonant limit of the parent model in the
present window; a paramagnetic-only truncation is not.

\section{Cavity-wave-vector control}
\label{app:sec4_k0}
The strip calculations in Sec.~\ref{sec:spectrum} use $k_0\ell_B=0.5$ as a
numerical working point at which the momentum ladder
$k\leftrightarrow k\pm 2k_0$ is spectrally resolved.  Deeply subwavelength
terahertz cavities typically realize $k_0\ell_B\ll 1$, where the
real-space separation $\Delta X=2k_0\ell_B^2$ between hybridized sectors is
small and the reconstruction is weak.  To separate a visibility choice from
a claim about present devices, we repeat the spectral analysis at fixed
$\lcav=0.2$ and $r_\omega=5$ for $k_0\ell_B=0.1$ and $0.5$.

\begin{figure}[!ht]
 \figcol{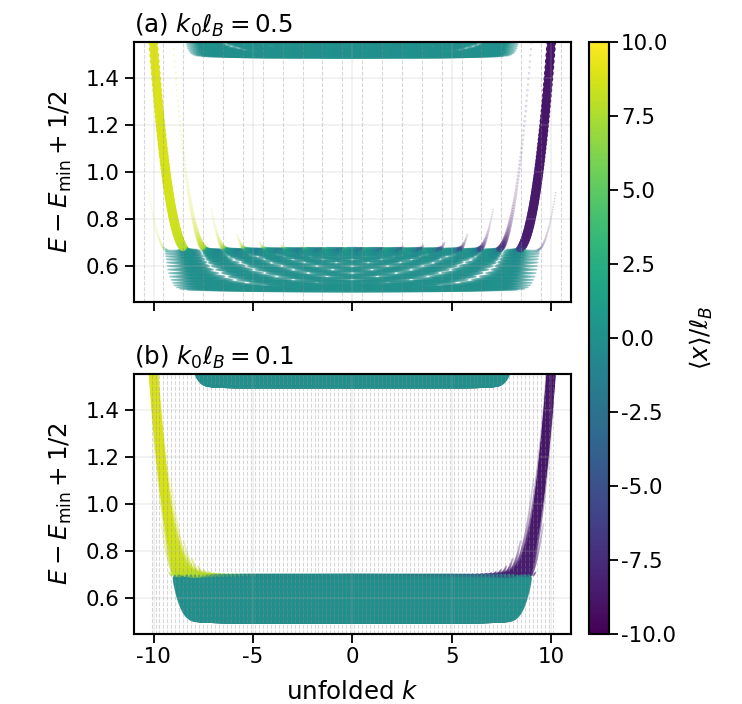}
 \caption{Cavity-wave-vector control of the momentum ladder at
 $\lcav=0.2$, $r_\omega=5$, and $L_x=20\ell_B$.
 Unfolded spectral weight at (a)~the working
 value $k_0\ell_B=0.5$ used in the main text and (b)~$k_0\ell_B=0.1$.
 Point size encodes the Fourier-sector weight and color encodes
 $\langle x\rangle/\ell_B$.
 Dashed vertical lines mark $2k_0$ zone boundaries.
 At $k_0\ell_B=0.1$ the $2k_0$ copies collapse onto one
 Landau-level branch and the retained subspace ends at $1.55$.}
 \label{fig:app_k0}
\end{figure}

Figure~\ref{fig:app_k0} compares the unfolded spectral weight at the two
wave vectors.  At $k_0\ell_B=0.1$ the $2k_0$ replicas overlap, and the
unfolded weight collapses onto a single Landau-level branch of the strip.
At $k_0\ell_B=0.5$ the copies separate by $\Delta k=2k_0\ell_B=1$, and
several Fourier sectors are resolved.
The comparison therefore has a definite reading.  The main-text
reconstruction is the spectrally resolved finite-$k_0$ limit of the same
Hamiltonian.  It is not the deeply subwavelength device limit, where the
ladder hybridization remains weak.

\section{Finite-width Hall plateaus}
\label{app:lx}
\begin{figure}[!ht]
 \figcol{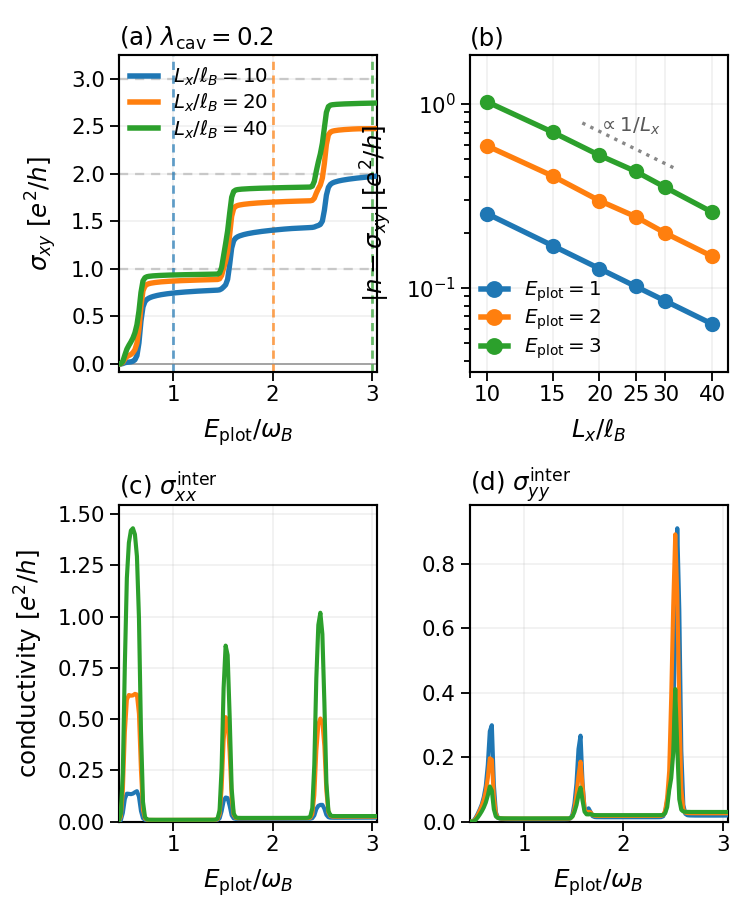}
 \caption{Finite-width Kubo response at $\lcav=0.2$ and $k_0\ell_B=0.5$,
 using the band-minimum-aligned energy
 $E_{\rm plot}=E_F-E_{\min}(L_x)+\wB/2$.
 (a)~Hall conductivity for $L_x/\ell_B=10,20,40$; dashed horizontals mark
 the integers and colored verticals the gap-center samples of (b).
 (b)~Gap-center residuals $|n-\sigma_{xy}|$ at
 $E_{\rm plot}/\wB=1,2,3$ versus $L_x$ on logarithmic axes;
 the dotted guide is $\propto 1/L_x$.
 (c),(d)~Transverse and along-edge interband longitudinal responses.
 As the strip widens the residuals fall as $1/L_x$,
 while $\sigma_{xx}^{\mathrm{inter}}$ and
 $\sigma_{yy}^{\mathrm{inter}}$ remain small in the gaps, with peaks
 at the Hall steps.}
 \label{fig:lx}
\end{figure}

Figure~\ref{fig:lx} shows that the noninteger Hall plateaus of
Fig.~\ref{fig:conductivity} at $L_x=20\ell_B$ are a finite-width effect,
while the cavity-induced step displacements survive as the strip is
widened.

At $\lcav=0.2$ the Hall residuals $|n-\sigma_{xy}|$ in
Fig.~\ref{fig:lx}(b) fall as $1/L_x$.
For $E_{\rm plot}/\wB=1,2,3$ they decrease from $0.254$, $0.593$,
and $1.030$ at $L_x=10\ell_B$ to $0.064$, $0.148$, and $0.258$
at $40\ell_B$, passing $0.127$, $0.298$, and $0.526$ at the working
width $L_x=20\ell_B$.
The added samples at $15$, $25$, and $30\,\ell_B$ lie on the same
$1/L_x$ trend.
Residual deviations at the working width $L_x=20\ell_B$ are an $O(10\%)$
strip-width effect, not a $10^{-3}$ artifact of $\eta$.
The same $1/L_x$ approach toward the integers is present at $\lcav=0$;
the cavity scan stops at $40\ell_B$ and establishes the trend rather
than a thermodynamic-limit extrapolation.

Figures~\ref{fig:lx}(c) and \ref{fig:lx}(d) show that
$\sigma_{xx}^{\mathrm{inter}}$ and $\sigma_{yy}^{\mathrm{inter}}$
remain small compared with $e^2/h$ inside the gaps, with the along-edge
interband peaks sitting at the Hall steps.
That size dependence supports reading the noninteger Hall plateaus of
Fig.~\ref{fig:conductivity} as finite-width artifacts.

\FloatBarrier
\bibliography{references}

\begin{thebibliography}{29}%
\makeatletter
\providecommand \@ifxundefined [1]{%
 \@ifx{#1\undefined}
}%
\providecommand \@ifnum [1]{%
 \ifnum #1\expandafter \@firstoftwo
 \else \expandafter \@secondoftwo
 \fi
}%
\providecommand \@ifx [1]{%
 \ifx #1\expandafter \@firstoftwo
 \else \expandafter \@secondoftwo
 \fi
}%
\providecommand \natexlab [1]{#1}%
\providecommand \enquote  [1]{``#1''}%
\providecommand \bibnamefont  [1]{#1}%
\providecommand \bibfnamefont [1]{#1}%
\providecommand \citenamefont [1]{#1}%
\providecommand \href@noop [0]{\@secondoftwo}%
\providecommand \href [0]{\begingroup \@sanitize@url \@href}%
\providecommand \@href[1]{\@@startlink{#1}\@@href}%
\providecommand \@@href[1]{\endgroup#1\@@endlink}%
\providecommand \@sanitize@url [0]{\catcode `\\12\catcode `\$12\catcode `\&12\catcode `\#12\catcode `\^12\catcode `\_12\catcode `\%12\relax}%
\providecommand \@@startlink[1]{}%
\providecommand \@@endlink[0]{}%
\providecommand \url  [0]{\begingroup\@sanitize@url \@url }%
\providecommand \@url [1]{\endgroup\@href {#1}{\urlprefix }}%
\providecommand \urlprefix  [0]{URL }%
\providecommand \Eprint [0]{\href }%
\providecommand \doibase [0]{https://doi.org/}%
\providecommand \selectlanguage [0]{\@gobble}%
\providecommand \bibinfo  [0]{\@secondoftwo}%
\providecommand \bibfield  [0]{\@secondoftwo}%
\providecommand \translation [1]{[#1]}%
\providecommand \BibitemOpen [0]{}%
\providecommand \bibitemStop [0]{}%
\providecommand \bibitemNoStop [0]{.\EOS\space}%
\providecommand \EOS [0]{\spacefactor3000\relax}%
\providecommand \BibitemShut  [1]{\csname bibitem#1\endcsname}%
\let\auto@bib@innerbib\@empty
\bibitem [{\citenamefont {von Klitzing}\ \emph {et~al.}(1980)\citenamefont {von Klitzing}, \citenamefont {Dorda},\ and\ \citenamefont {Pepper}}]{vonKlitzing1980}%
  \BibitemOpen
  \bibfield  {author} {\bibinfo {author} {\bibfnamefont {K.}~\bibnamefont {von Klitzing}}, \bibinfo {author} {\bibfnamefont {G.}~\bibnamefont {Dorda}},\ and\ \bibinfo {author} {\bibfnamefont {M.}~\bibnamefont {Pepper}},\ }\bibfield  {title} {\bibinfo {title} {New method for high-accuracy determination of the fine-structure constant based on quantized {H}all resistance},\ }\href {https://doi.org/10.1103/PhysRevLett.45.494} {\bibfield  {journal} {\bibinfo  {journal} {Phys. Rev. Lett.}\ }\textbf {\bibinfo {volume} {45}},\ \bibinfo {pages} {494} (\bibinfo {year} {1980})}\BibitemShut {NoStop}%
\bibitem [{\citenamefont {Laughlin}(1981)}]{Laughlin1981}%
  \BibitemOpen
  \bibfield  {author} {\bibinfo {author} {\bibfnamefont {R.~B.}\ \bibnamefont {Laughlin}},\ }\bibfield  {title} {\bibinfo {title} {Quantized {H}all conductivity in two dimensions},\ }\href {https://doi.org/10.1103/PhysRevB.23.5632} {\bibfield  {journal} {\bibinfo  {journal} {Phys. Rev. B}\ }\textbf {\bibinfo {volume} {23}},\ \bibinfo {pages} {5632} (\bibinfo {year} {1981})}\BibitemShut {NoStop}%
\bibitem [{\citenamefont {B{\"u}ttiker}(1988)}]{Buttiker1988}%
  \BibitemOpen
  \bibfield  {author} {\bibinfo {author} {\bibfnamefont {M.}~\bibnamefont {B{\"u}ttiker}},\ }\bibfield  {title} {\bibinfo {title} {Absence of backscattering in the quantum {H}all effect in multiprobe conductors},\ }\href {https://doi.org/10.1103/PhysRevB.38.9375} {\bibfield  {journal} {\bibinfo  {journal} {Phys. Rev. B}\ }\textbf {\bibinfo {volume} {38}},\ \bibinfo {pages} {9375} (\bibinfo {year} {1988})}\BibitemShut {NoStop}%
\bibitem [{\citenamefont {Pruisken}(1984)}]{Pruisken1984}%
  \BibitemOpen
  \bibfield  {author} {\bibinfo {author} {\bibfnamefont {A.~M.~M.}\ \bibnamefont {Pruisken}},\ }\bibfield  {title} {\bibinfo {title} {On localization in the theory of the quantized {H}all effect: A two-dimensional realization of the $\theta$-vacuum},\ }\href {https://doi.org/10.1016/0550-3213(84)90101-9} {\bibfield  {journal} {\bibinfo  {journal} {Nuclear Phys. B}\ }\textbf {\bibinfo {volume} {235}},\ \bibinfo {pages} {1} (\bibinfo {year} {1984})}\BibitemShut {NoStop}%
\bibitem [{\citenamefont {Khmelnitskii}(1984)}]{Khmelnitskii1984}%
  \BibitemOpen
  \bibfield  {author} {\bibinfo {author} {\bibfnamefont {D.~E.}\ \bibnamefont {Khmelnitskii}},\ }\bibfield  {title} {\bibinfo {title} {Quantum {H}all effect and additional oscillations of conductivity in weak magnetic fields},\ }\href {https://doi.org/10.1016/0375-9601(84)90976-6} {\bibfield  {journal} {\bibinfo  {journal} {Phys. Lett. A}\ }\textbf {\bibinfo {volume} {106}},\ \bibinfo {pages} {182} (\bibinfo {year} {1984})}\BibitemShut {NoStop}%
\bibitem [{\citenamefont {Wei}\ \emph {et~al.}(1988)\citenamefont {Wei}, \citenamefont {Tsui}, \citenamefont {Paalanen},\ and\ \citenamefont {Pruisken}}]{Wei1988}%
  \BibitemOpen
  \bibfield  {author} {\bibinfo {author} {\bibfnamefont {H.~P.}\ \bibnamefont {Wei}}, \bibinfo {author} {\bibfnamefont {D.~C.}\ \bibnamefont {Tsui}}, \bibinfo {author} {\bibfnamefont {M.~A.}\ \bibnamefont {Paalanen}},\ and\ \bibinfo {author} {\bibfnamefont {A.~M.~M.}\ \bibnamefont {Pruisken}},\ }\bibfield  {title} {\bibinfo {title} {Experiments on delocalization and universality in the integral quantum {H}all effect},\ }\href {https://doi.org/10.1103/PhysRevLett.61.1294} {\bibfield  {journal} {\bibinfo  {journal} {Phys. Rev. Lett.}\ }\textbf {\bibinfo {volume} {61}},\ \bibinfo {pages} {1294} (\bibinfo {year} {1988})}\BibitemShut {NoStop}%
\bibitem [{\citenamefont {Shahar}\ \emph {et~al.}(1995)\citenamefont {Shahar}, \citenamefont {Tsui}, \citenamefont {Shayegan}, \citenamefont {Bhatt},\ and\ \citenamefont {Cunningham}}]{Shahar1995}%
  \BibitemOpen
  \bibfield  {author} {\bibinfo {author} {\bibfnamefont {D.}~\bibnamefont {Shahar}}, \bibinfo {author} {\bibfnamefont {D.~C.}\ \bibnamefont {Tsui}}, \bibinfo {author} {\bibfnamefont {M.}~\bibnamefont {Shayegan}}, \bibinfo {author} {\bibfnamefont {R.~N.}\ \bibnamefont {Bhatt}},\ and\ \bibinfo {author} {\bibfnamefont {J.~E.}\ \bibnamefont {Cunningham}},\ }\bibfield  {title} {\bibinfo {title} {Universal conductivity at the quantum {H}all liquid to insulator transition},\ }\href {https://doi.org/10.1103/PhysRevLett.74.4511} {\bibfield  {journal} {\bibinfo  {journal} {Phys. Rev. Lett.}\ }\textbf {\bibinfo {volume} {74}},\ \bibinfo {pages} {4511} (\bibinfo {year} {1995})}\BibitemShut {NoStop}%
\bibitem [{\citenamefont {Ostrovsky}\ \emph {et~al.}(2007)\citenamefont {Ostrovsky}, \citenamefont {Gornyi},\ and\ \citenamefont {Mirlin}}]{Ostrovsky2007PRL}%
  \BibitemOpen
  \bibfield  {author} {\bibinfo {author} {\bibfnamefont {P.~M.}\ \bibnamefont {Ostrovsky}}, \bibinfo {author} {\bibfnamefont {I.~V.}\ \bibnamefont {Gornyi}},\ and\ \bibinfo {author} {\bibfnamefont {A.~D.}\ \bibnamefont {Mirlin}},\ }\bibfield  {title} {\bibinfo {title} {Quantum criticality and minimal conductivity in graphene with long-range disorder},\ }\href {https://doi.org/10.1103/PhysRevLett.98.256801} {\bibfield  {journal} {\bibinfo  {journal} {Phys. Rev. Lett.}\ }\textbf {\bibinfo {volume} {98}},\ \bibinfo {pages} {256801} (\bibinfo {year} {2007})}\BibitemShut {NoStop}%
\bibitem [{\citenamefont {Altland}\ \emph {et~al.}(2014)\citenamefont {Altland}, \citenamefont {Bagrets}, \citenamefont {Fritz}, \citenamefont {Kamenev},\ and\ \citenamefont {Schmiedt}}]{Altland2014}%
  \BibitemOpen
  \bibfield  {author} {\bibinfo {author} {\bibfnamefont {A.}~\bibnamefont {Altland}}, \bibinfo {author} {\bibfnamefont {D.}~\bibnamefont {Bagrets}}, \bibinfo {author} {\bibfnamefont {L.}~\bibnamefont {Fritz}}, \bibinfo {author} {\bibfnamefont {A.}~\bibnamefont {Kamenev}},\ and\ \bibinfo {author} {\bibfnamefont {H.}~\bibnamefont {Schmiedt}},\ }\bibfield  {title} {\bibinfo {title} {Quantum criticality of quasi-one-dimensional topological {A}nderson insulators},\ }\href {https://doi.org/10.1103/PhysRevLett.112.206602} {\bibfield  {journal} {\bibinfo  {journal} {Phys. Rev. Lett.}\ }\textbf {\bibinfo {volume} {112}},\ \bibinfo {pages} {206602} (\bibinfo {year} {2014})}\BibitemShut {NoStop}%
\bibitem [{\citenamefont {Groth}\ \emph {et~al.}(2009)\citenamefont {Groth}, \citenamefont {Wimmer}, \citenamefont {Akhmerov}, \citenamefont {Tworzyd{\l}o},\ and\ \citenamefont {Beenakker}}]{Groth2009}%
  \BibitemOpen
  \bibfield  {author} {\bibinfo {author} {\bibfnamefont {C.~W.}\ \bibnamefont {Groth}}, \bibinfo {author} {\bibfnamefont {M.}~\bibnamefont {Wimmer}}, \bibinfo {author} {\bibfnamefont {A.~R.}\ \bibnamefont {Akhmerov}}, \bibinfo {author} {\bibfnamefont {J.}~\bibnamefont {Tworzyd{\l}o}},\ and\ \bibinfo {author} {\bibfnamefont {C.~W.~J.}\ \bibnamefont {Beenakker}},\ }\bibfield  {title} {\bibinfo {title} {Theory of the topological {A}nderson insulator},\ }\href {https://doi.org/10.1103/PhysRevLett.103.196805} {\bibfield  {journal} {\bibinfo  {journal} {Phys. Rev. Lett.}\ }\textbf {\bibinfo {volume} {103}},\ \bibinfo {pages} {196805} (\bibinfo {year} {2009})}\BibitemShut {NoStop}%
\bibitem [{\citenamefont {Li}\ \emph {et~al.}(2009)\citenamefont {Li}, \citenamefont {Chu}, \citenamefont {Jain},\ and\ \citenamefont {Shen}}]{Li2009}%
  \BibitemOpen
  \bibfield  {author} {\bibinfo {author} {\bibfnamefont {J.}~\bibnamefont {Li}}, \bibinfo {author} {\bibfnamefont {R.-L.}\ \bibnamefont {Chu}}, \bibinfo {author} {\bibfnamefont {J.~K.}\ \bibnamefont {Jain}},\ and\ \bibinfo {author} {\bibfnamefont {S.-Q.}\ \bibnamefont {Shen}},\ }\bibfield  {title} {\bibinfo {title} {Topological {A}nderson insulator},\ }\href {https://doi.org/10.1103/PhysRevLett.102.136806} {\bibfield  {journal} {\bibinfo  {journal} {Phys. Rev. Lett.}\ }\textbf {\bibinfo {volume} {102}},\ \bibinfo {pages} {136806} (\bibinfo {year} {2009})}\BibitemShut {NoStop}%
\bibitem [{\citenamefont {van Nieuwenburg}\ \emph {et~al.}(2012)\citenamefont {van Nieuwenburg}, \citenamefont {Edge}, \citenamefont {Dahlhaus}, \citenamefont {Tworzyd{\l}o},\ and\ \citenamefont {Beenakker}}]{vanNieuwenburg2012}%
  \BibitemOpen
  \bibfield  {author} {\bibinfo {author} {\bibfnamefont {E.~P.~L.}\ \bibnamefont {van Nieuwenburg}}, \bibinfo {author} {\bibfnamefont {J.~M.}\ \bibnamefont {Edge}}, \bibinfo {author} {\bibfnamefont {J.~P.}\ \bibnamefont {Dahlhaus}}, \bibinfo {author} {\bibfnamefont {J.}~\bibnamefont {Tworzyd{\l}o}},\ and\ \bibinfo {author} {\bibfnamefont {C.~W.~J.}\ \bibnamefont {Beenakker}},\ }\bibfield  {title} {\bibinfo {title} {Metal-topological-insulator transition in the quantum kicked rotator with {$\mathbb{Z}_2$} symmetry},\ }\href {https://doi.org/10.1103/PhysRevB.85.165131} {\bibfield  {journal} {\bibinfo  {journal} {Phys. Rev. B}\ }\textbf {\bibinfo {volume} {85}},\ \bibinfo {pages} {165131} (\bibinfo {year} {2012})}\BibitemShut {NoStop}%
\bibitem [{\citenamefont {Edge}\ \emph {et~al.}(2012)\citenamefont {Edge}, \citenamefont {Tworzyd{\l}o},\ and\ \citenamefont {Beenakker}}]{Edge2012}%
  \BibitemOpen
  \bibfield  {author} {\bibinfo {author} {\bibfnamefont {J.~M.}\ \bibnamefont {Edge}}, \bibinfo {author} {\bibfnamefont {J.}~\bibnamefont {Tworzyd{\l}o}},\ and\ \bibinfo {author} {\bibfnamefont {C.~W.~J.}\ \bibnamefont {Beenakker}},\ }\bibfield  {title} {\bibinfo {title} {Metallic phase of the quantum {H}all effect in four-dimensional space},\ }\href {https://doi.org/10.1103/PhysRevLett.109.135701} {\bibfield  {journal} {\bibinfo  {journal} {Phys. Rev. Lett.}\ }\textbf {\bibinfo {volume} {109}},\ \bibinfo {pages} {135701} (\bibinfo {year} {2012})}\BibitemShut {NoStop}%
\bibitem [{\citenamefont {Chen}\ and\ \citenamefont {Tian}(2014)}]{Chen2014}%
  \BibitemOpen
  \bibfield  {author} {\bibinfo {author} {\bibfnamefont {Y.}~\bibnamefont {Chen}}\ and\ \bibinfo {author} {\bibfnamefont {C.}~\bibnamefont {Tian}},\ }\bibfield  {title} {\bibinfo {title} {Planck's quantum-driven integer quantum {H}all effect in chaos},\ }\href {https://doi.org/10.1103/PhysRevLett.113.216802} {\bibfield  {journal} {\bibinfo  {journal} {Phys. Rev. Lett.}\ }\textbf {\bibinfo {volume} {113}},\ \bibinfo {pages} {216802} (\bibinfo {year} {2014})}\BibitemShut {NoStop}%
\bibitem [{\citenamefont {Tian}\ \emph {et~al.}(2016)\citenamefont {Tian}, \citenamefont {Chen},\ and\ \citenamefont {Wang}}]{Tian2016}%
  \BibitemOpen
  \bibfield  {author} {\bibinfo {author} {\bibfnamefont {C.}~\bibnamefont {Tian}}, \bibinfo {author} {\bibfnamefont {Y.}~\bibnamefont {Chen}},\ and\ \bibinfo {author} {\bibfnamefont {J.}~\bibnamefont {Wang}},\ }\bibfield  {title} {\bibinfo {title} {Emergence of integer quantum {H}all effect from chaos},\ }\href {https://doi.org/10.1103/PhysRevB.93.075403} {\bibfield  {journal} {\bibinfo  {journal} {Phys. Rev. B}\ }\textbf {\bibinfo {volume} {93}},\ \bibinfo {pages} {075403} (\bibinfo {year} {2016})}\BibitemShut {NoStop}%
\bibitem [{\citenamefont {Titum}\ \emph {et~al.}(2015)\citenamefont {Titum}, \citenamefont {Lindner}, \citenamefont {Rechtsman},\ and\ \citenamefont {Refael}}]{Titum2015}%
  \BibitemOpen
  \bibfield  {author} {\bibinfo {author} {\bibfnamefont {P.}~\bibnamefont {Titum}}, \bibinfo {author} {\bibfnamefont {N.~H.}\ \bibnamefont {Lindner}}, \bibinfo {author} {\bibfnamefont {M.~C.}\ \bibnamefont {Rechtsman}},\ and\ \bibinfo {author} {\bibfnamefont {G.}~\bibnamefont {Refael}},\ }\bibfield  {title} {\bibinfo {title} {Disorder-induced {F}loquet topological insulators},\ }\href {https://doi.org/10.1103/PhysRevLett.114.056801} {\bibfield  {journal} {\bibinfo  {journal} {Phys. Rev. Lett.}\ }\textbf {\bibinfo {volume} {114}},\ \bibinfo {pages} {056801} (\bibinfo {year} {2015})}\BibitemShut {NoStop}%
\bibitem [{\citenamefont {Titum}\ \emph {et~al.}(2016)\citenamefont {Titum}, \citenamefont {Berg}, \citenamefont {Rudner}, \citenamefont {Refael},\ and\ \citenamefont {Lindner}}]{Titum2016}%
  \BibitemOpen
  \bibfield  {author} {\bibinfo {author} {\bibfnamefont {P.}~\bibnamefont {Titum}}, \bibinfo {author} {\bibfnamefont {E.}~\bibnamefont {Berg}}, \bibinfo {author} {\bibfnamefont {M.~S.}\ \bibnamefont {Rudner}}, \bibinfo {author} {\bibfnamefont {G.}~\bibnamefont {Refael}},\ and\ \bibinfo {author} {\bibfnamefont {N.~H.}\ \bibnamefont {Lindner}},\ }\bibfield  {title} {\bibinfo {title} {Anomalous {F}loquet-{A}nderson insulator as a nonadiabatic quantized charge pump},\ }\href {https://doi.org/10.1103/PhysRevX.6.021013} {\bibfield  {journal} {\bibinfo  {journal} {Phys. Rev. X}\ }\textbf {\bibinfo {volume} {6}},\ \bibinfo {pages} {021013} (\bibinfo {year} {2016})}\BibitemShut {NoStop}%
\bibitem [{\citenamefont {Kim}\ \emph {et~al.}(2020)\citenamefont {Kim}, \citenamefont {Bagrets}, \citenamefont {Micklitz},\ and\ \citenamefont {Altland}}]{Kim2020AFAI}%
  \BibitemOpen
  \bibfield  {author} {\bibinfo {author} {\bibfnamefont {K.~W.}\ \bibnamefont {Kim}}, \bibinfo {author} {\bibfnamefont {D.}~\bibnamefont {Bagrets}}, \bibinfo {author} {\bibfnamefont {T.}~\bibnamefont {Micklitz}},\ and\ \bibinfo {author} {\bibfnamefont {A.}~\bibnamefont {Altland}},\ }\bibfield  {title} {\bibinfo {title} {Quantum {H}all criticality in {F}loquet topological insulators},\ }\href {https://doi.org/10.1103/PhysRevB.101.165401} {\bibfield  {journal} {\bibinfo  {journal} {Phys. Rev. B}\ }\textbf {\bibinfo {volume} {101}},\ \bibinfo {pages} {165401} (\bibinfo {year} {2020})}\BibitemShut {NoStop}%
\bibitem [{\citenamefont {Hagenm{\"u}ller}\ \emph {et~al.}(2010)\citenamefont {Hagenm{\"u}ller}, \citenamefont {Liberato},\ and\ \citenamefont {Ciuti}}]{Hagenmuller2010}%
  \BibitemOpen
  \bibfield  {author} {\bibinfo {author} {\bibfnamefont {D.}~\bibnamefont {Hagenm{\"u}ller}}, \bibinfo {author} {\bibfnamefont {S.~D.}\ \bibnamefont {Liberato}},\ and\ \bibinfo {author} {\bibfnamefont {C.}~\bibnamefont {Ciuti}},\ }\bibfield  {title} {\bibinfo {title} {Ultrastrong coupling between a cavity resonator and the cyclotron transition of a two-dimensional electron gas in the case of an integer filling factor},\ }\href {https://doi.org/10.1103/PhysRevB.81.235303} {\bibfield  {journal} {\bibinfo  {journal} {Phys. Rev. B}\ }\textbf {\bibinfo {volume} {81}},\ \bibinfo {pages} {235303} (\bibinfo {year} {2010})}\BibitemShut {NoStop}%
\bibitem [{\citenamefont {Li}\ \emph {et~al.}(2018)\citenamefont {Li}, \citenamefont {Bamba}, \citenamefont {Zhang}, \citenamefont {Fallahi}, \citenamefont {Gardner}, \citenamefont {Gao}, \citenamefont {Lou}, \citenamefont {Yoshioka}, \citenamefont {Manfra},\ and\ \citenamefont {Kono}}]{Li2018}%
  \BibitemOpen
  \bibfield  {author} {\bibinfo {author} {\bibfnamefont {X.}~\bibnamefont {Li}}, \bibinfo {author} {\bibfnamefont {M.}~\bibnamefont {Bamba}}, \bibinfo {author} {\bibfnamefont {Q.}~\bibnamefont {Zhang}}, \bibinfo {author} {\bibfnamefont {S.}~\bibnamefont {Fallahi}}, \bibinfo {author} {\bibfnamefont {G.~C.}\ \bibnamefont {Gardner}}, \bibinfo {author} {\bibfnamefont {W.}~\bibnamefont {Gao}}, \bibinfo {author} {\bibfnamefont {M.}~\bibnamefont {Lou}}, \bibinfo {author} {\bibfnamefont {K.}~\bibnamefont {Yoshioka}}, \bibinfo {author} {\bibfnamefont {M.~J.}\ \bibnamefont {Manfra}},\ and\ \bibinfo {author} {\bibfnamefont {J.}~\bibnamefont {Kono}},\ }\bibfield  {title} {\bibinfo {title} {Vacuum {B}loch--{S}iegert shift in {L}andau polaritons with ultra-high cooperativity},\ }\href {https://doi.org/10.1038/s41566-018-0153-0} {\bibfield  {journal} {\bibinfo  {journal} {Nat. Photonics}\ }\textbf {\bibinfo {volume} {12}},\ \bibinfo {pages} {324} (\bibinfo {year} {2018})}\BibitemShut {NoStop}%
\bibitem [{\citenamefont {Paravicini-Bagliani}\ \emph {et~al.}(2019)\citenamefont {Paravicini-Bagliani}, \citenamefont {Appugliese}, \citenamefont {Richter}, \citenamefont {Valmorra}, \citenamefont {Keller}, \citenamefont {Beck}, \citenamefont {Bartolo}, \citenamefont {R{\"o}ssler}, \citenamefont {Ihn}, \citenamefont {Ensslin}, \citenamefont {Ciuti}, \citenamefont {Scalari},\ and\ \citenamefont {Faist}}]{ParaviciniBagliani2019}%
  \BibitemOpen
  \bibfield  {author} {\bibinfo {author} {\bibfnamefont {G.~L.}\ \bibnamefont {Paravicini-Bagliani}}, \bibinfo {author} {\bibfnamefont {F.}~\bibnamefont {Appugliese}}, \bibinfo {author} {\bibfnamefont {E.}~\bibnamefont {Richter}}, \bibinfo {author} {\bibfnamefont {F.}~\bibnamefont {Valmorra}}, \bibinfo {author} {\bibfnamefont {J.}~\bibnamefont {Keller}}, \bibinfo {author} {\bibfnamefont {M.}~\bibnamefont {Beck}}, \bibinfo {author} {\bibfnamefont {N.}~\bibnamefont {Bartolo}}, \bibinfo {author} {\bibfnamefont {C.}~\bibnamefont {R{\"o}ssler}}, \bibinfo {author} {\bibfnamefont {T.}~\bibnamefont {Ihn}}, \bibinfo {author} {\bibfnamefont {K.}~\bibnamefont {Ensslin}}, \bibinfo {author} {\bibfnamefont {C.}~\bibnamefont {Ciuti}}, \bibinfo {author} {\bibfnamefont {G.}~\bibnamefont {Scalari}},\ and\ \bibinfo {author} {\bibfnamefont {J.}~\bibnamefont {Faist}},\ }\bibfield  {title} {\bibinfo {title} {Magneto-transport controlled by {L}andau polariton states},\ }\href {https://doi.org/10.1038/s41567-018-0346-y} {\bibfield  {journal} {\bibinfo  {journal} {Nat. Phys.}\ }\textbf {\bibinfo {volume} {15}},\ \bibinfo {pages} {186} (\bibinfo {year} {2019})}\BibitemShut {NoStop}%
\bibitem [{\citenamefont {Appugliese}\ \emph {et~al.}(2022)\citenamefont {Appugliese}, \citenamefont {Enkner}, \citenamefont {Paravicini-Bagliani}, \citenamefont {Beck}, \citenamefont {Reichl}, \citenamefont {Wegscheider}, \citenamefont {Scalari}, \citenamefont {Ciuti},\ and\ \citenamefont {Faist}}]{Appugliese2022}%
  \BibitemOpen
  \bibfield  {author} {\bibinfo {author} {\bibfnamefont {F.}~\bibnamefont {Appugliese}}, \bibinfo {author} {\bibfnamefont {J.}~\bibnamefont {Enkner}}, \bibinfo {author} {\bibfnamefont {G.~L.}\ \bibnamefont {Paravicini-Bagliani}}, \bibinfo {author} {\bibfnamefont {M.}~\bibnamefont {Beck}}, \bibinfo {author} {\bibfnamefont {C.}~\bibnamefont {Reichl}}, \bibinfo {author} {\bibfnamefont {W.}~\bibnamefont {Wegscheider}}, \bibinfo {author} {\bibfnamefont {G.}~\bibnamefont {Scalari}}, \bibinfo {author} {\bibfnamefont {C.}~\bibnamefont {Ciuti}},\ and\ \bibinfo {author} {\bibfnamefont {J.}~\bibnamefont {Faist}},\ }\bibfield  {title} {\bibinfo {title} {Breakdown of the topological protection by cavity vacuum fields in the integer quantum {H}all effect},\ }\href {https://doi.org/10.1126/science.abl5818} {\bibfield  {journal} {\bibinfo  {journal} {Science}\ }\textbf {\bibinfo {volume} {375}},\ \bibinfo {pages} {1030} (\bibinfo {year} {2022})}\BibitemShut {NoStop}%
\bibitem [{\citenamefont {Enkner}\ \emph {et~al.}(2025)\citenamefont {Enkner}, \citenamefont {Graziotto}, \citenamefont {Bori{\c c}i}, \citenamefont {Appugliese}, \citenamefont {Reichl}, \citenamefont {Scalari}, \citenamefont {Regnault}, \citenamefont {Wegscheider}, \citenamefont {Ciuti},\ and\ \citenamefont {Faist}}]{Enkner2025}%
  \BibitemOpen
  \bibfield  {author} {\bibinfo {author} {\bibfnamefont {J.}~\bibnamefont {Enkner}}, \bibinfo {author} {\bibfnamefont {L.}~\bibnamefont {Graziotto}}, \bibinfo {author} {\bibfnamefont {D.}~\bibnamefont {Bori{\c c}i}}, \bibinfo {author} {\bibfnamefont {F.}~\bibnamefont {Appugliese}}, \bibinfo {author} {\bibfnamefont {C.}~\bibnamefont {Reichl}}, \bibinfo {author} {\bibfnamefont {G.}~\bibnamefont {Scalari}}, \bibinfo {author} {\bibfnamefont {N.}~\bibnamefont {Regnault}}, \bibinfo {author} {\bibfnamefont {W.}~\bibnamefont {Wegscheider}}, \bibinfo {author} {\bibfnamefont {C.}~\bibnamefont {Ciuti}},\ and\ \bibinfo {author} {\bibfnamefont {J.}~\bibnamefont {Faist}},\ }\bibfield  {title} {\bibinfo {title} {Tunable vacuum-field control of fractional and integer quantum {H}all phases},\ }\href {https://doi.org/10.1038/s41586-025-08894-3} {\bibfield  {journal} {\bibinfo  {journal} {Nature}\ }\textbf {\bibinfo {volume} {641}},\ \bibinfo {pages} {884} (\bibinfo {year} {2025})}\BibitemShut {NoStop}%
\bibitem [{\citenamefont {Graziotto}\ \emph {et~al.}(2026)\citenamefont {Graziotto}, \citenamefont {Enkner}, \citenamefont {Chattopadhyay}, \citenamefont {Curtis}, \citenamefont {Koskas}, \citenamefont {Reichl}, \citenamefont {Wegscheider}, \citenamefont {Scalari}, \citenamefont {Demler},\ and\ \citenamefont {Faist}}]{Graziotto2026}%
  \BibitemOpen
  \bibfield  {author} {\bibinfo {author} {\bibfnamefont {L.}~\bibnamefont {Graziotto}}, \bibinfo {author} {\bibfnamefont {J.}~\bibnamefont {Enkner}}, \bibinfo {author} {\bibfnamefont {S.}~\bibnamefont {Chattopadhyay}}, \bibinfo {author} {\bibfnamefont {J.~B.}\ \bibnamefont {Curtis}}, \bibinfo {author} {\bibfnamefont {E.}~\bibnamefont {Koskas}}, \bibinfo {author} {\bibfnamefont {C.}~\bibnamefont {Reichl}}, \bibinfo {author} {\bibfnamefont {W.}~\bibnamefont {Wegscheider}}, \bibinfo {author} {\bibfnamefont {G.}~\bibnamefont {Scalari}}, \bibinfo {author} {\bibfnamefont {E.}~\bibnamefont {Demler}},\ and\ \bibinfo {author} {\bibfnamefont {J.}~\bibnamefont {Faist}},\ }\bibfield  {title} {\bibinfo {title} {Cavity quantum electrodynamics control of quantum {H}all stripes},\ }\href {https://doi.org/10.1038/s41567-026-03287-3} {\bibfield  {journal} {\bibinfo  {journal} {Nat. Phys.}\ }\textbf {\bibinfo {volume} {22}},\ \bibinfo {pages} {1094} (\bibinfo {year} {2026})}\BibitemShut {NoStop}%
\bibitem [{\citenamefont {Bartolo}\ and\ \citenamefont {Ciuti}(2018)}]{Bartolo2018}%
  \BibitemOpen
  \bibfield  {author} {\bibinfo {author} {\bibfnamefont {N.}~\bibnamefont {Bartolo}}\ and\ \bibinfo {author} {\bibfnamefont {C.}~\bibnamefont {Ciuti}},\ }\bibfield  {title} {\bibinfo {title} {Vacuum-dressed cavity magnetotransport of a two-dimensional electron gas},\ }\href {https://doi.org/10.1103/PhysRevB.98.205301} {\bibfield  {journal} {\bibinfo  {journal} {Phys. Rev. B}\ }\textbf {\bibinfo {volume} {98}},\ \bibinfo {pages} {205301} (\bibinfo {year} {2018})}\BibitemShut {NoStop}%
\bibitem [{\citenamefont {Rokaj}\ \emph {et~al.}(2022)\citenamefont {Rokaj}, \citenamefont {Penz}, \citenamefont {Sentef}, \citenamefont {Ruggenthaler},\ and\ \citenamefont {Rubio}}]{Rokaj2022}%
  \BibitemOpen
  \bibfield  {author} {\bibinfo {author} {\bibfnamefont {V.}~\bibnamefont {Rokaj}}, \bibinfo {author} {\bibfnamefont {M.}~\bibnamefont {Penz}}, \bibinfo {author} {\bibfnamefont {M.~A.}\ \bibnamefont {Sentef}}, \bibinfo {author} {\bibfnamefont {M.}~\bibnamefont {Ruggenthaler}},\ and\ \bibinfo {author} {\bibfnamefont {A.}~\bibnamefont {Rubio}},\ }\bibfield  {title} {\bibinfo {title} {Polaritonic {H}ofstadter butterfly and cavity control of the quantized {H}all conductance},\ }\href {https://doi.org/10.1103/PhysRevB.105.205424} {\bibfield  {journal} {\bibinfo  {journal} {Phys. Rev. B}\ }\textbf {\bibinfo {volume} {105}},\ \bibinfo {pages} {205424} (\bibinfo {year} {2022})}\BibitemShut {NoStop}%
\bibitem [{\citenamefont {Bacciconi}\ \emph {et~al.}(2025)\citenamefont {Bacciconi}, \citenamefont {Xavier}, \citenamefont {Carusotto}, \citenamefont {Chanda},\ and\ \citenamefont {Dalmonte}}]{Bacciconi2025}%
  \BibitemOpen
  \bibfield  {author} {\bibinfo {author} {\bibfnamefont {Z.}~\bibnamefont {Bacciconi}}, \bibinfo {author} {\bibfnamefont {H.~B.}\ \bibnamefont {Xavier}}, \bibinfo {author} {\bibfnamefont {I.}~\bibnamefont {Carusotto}}, \bibinfo {author} {\bibfnamefont {T.}~\bibnamefont {Chanda}},\ and\ \bibinfo {author} {\bibfnamefont {M.}~\bibnamefont {Dalmonte}},\ }\bibfield  {title} {\bibinfo {title} {Theory of fractional quantum {H}all liquids coupled to quantum light and emergent graviton-polaritons},\ }\href {https://doi.org/10.1103/PhysRevX.15.021027} {\bibfield  {journal} {\bibinfo  {journal} {Phys. Rev. X}\ }\textbf {\bibinfo {volume} {15}},\ \bibinfo {pages} {021027} (\bibinfo {year} {2025})}\BibitemShut {NoStop}%
\bibitem [{\citenamefont {Ciuti}(2021)}]{Ciuti2021}%
  \BibitemOpen
  \bibfield  {author} {\bibinfo {author} {\bibfnamefont {C.}~\bibnamefont {Ciuti}},\ }\bibfield  {title} {\bibinfo {title} {Cavity-mediated electron hopping in disordered quantum {H}all systems},\ }\href {https://doi.org/10.1103/PhysRevB.104.155303} {\bibfield  {journal} {\bibinfo  {journal} {Phys. Rev. B}\ }\textbf {\bibinfo {volume} {104}},\ \bibinfo {pages} {155303} (\bibinfo {year} {2021})}\BibitemShut {NoStop}%
\bibitem [{\citenamefont {Fukui}\ \emph {et~al.}(2005)\citenamefont {Fukui}, \citenamefont {Hatsugai},\ and\ \citenamefont {Suzuki}}]{Fukui2005}%
  \BibitemOpen
  \bibfield  {author} {\bibinfo {author} {\bibfnamefont {T.}~\bibnamefont {Fukui}}, \bibinfo {author} {\bibfnamefont {Y.}~\bibnamefont {Hatsugai}},\ and\ \bibinfo {author} {\bibfnamefont {H.}~\bibnamefont {Suzuki}},\ }\bibfield  {title} {\bibinfo {title} {Chern numbers in discretized {B}rillouin zone: Efficient method of computing ({S}pin) {H}all conductances},\ }\href {https://doi.org/10.1143/JPSJ.74.1674} {\bibfield  {journal} {\bibinfo  {journal} {J. Phys. Soc. Jpn.}\ }\textbf {\bibinfo {volume} {74}},\ \bibinfo {pages} {1674} (\bibinfo {year} {2005})}\BibitemShut {NoStop}%
\end{thebibliography}%

\end{document}